\documentclass{article}
\usepackage{a4}
\usepackage[dvips]{color}
\usepackage[dvips]{graphicx}
\usepackage{amsmath}
\usepackage{amssymb}

\begin{document}
\begin{titlepage}
\begin{flushright}LMU-ETP-2003-01\end{flushright}
\vspace{0.5cm}
\begin{center}
{\Large\bf
A Cosmic Ray Measurement Facility\vspace{1ex}\\ for ATLAS Muon Chambers
}
\end{center}
\vspace{2cm}

\begin{center}
O.~Biebel$^{1}$,
M.~Binder$^{1}$,
M.~Boutemeur$^{1}$,
A.~Brandt$^{1}$,
J.~Dubbert$^{1}$,
G.~Duckeck$^{1}$,
J.~Elmsheuser$^{1}$,
F.~Fiedler$^{1}$,
R.~Hertenberger$^{1}$,
O.~Kortner$^{2}$,
T.~Nunnemann$^{1}$,
F.~Rauscher$^{1}$,
D.~Schaile$^{1}$,
P.~Schieferdecker$^{1}$,
A.~Staude$^{1}$,
W.~Stiller$^{2}$,
R.~Str\"ohmer$^{1}$, and
R.~V\'ertesi$^{1}$
\end{center}
\begin{center}
(1): Ludwig-Maximilians-Universit\"at M\"unchen,\\ 
Am Coulombwall 1, D-85748 Garching, Germany\\
(2): Max-Planck-Institut f\"ur Physik (Werner-Heisenberg-Institut),\\
F\"ohringer Ring 6, D-80805 M\"unchen, Germany
\end{center}
\vspace{2cm}
\begin{abstract}
Monitored Drift Tube (MDT) chambers will constitute the large majority
of precision detectors in the Muon Spectrometer of the ATLAS
experiment at the Large Hadron Collider at CERN.
For commissioning and calibration of MDT chambers, a Cosmic Ray
Measurement Facility is in operation at Munich University.  The
objectives of this facility are to test the chambers and on-chamber
electronics, to map the positions of the anode wires within the
chambers with the precision needed for standalone muon momentum
measurement in ATLAS, and to gain experience in the operation of the
chambers and on-line calibration procedures.

Until the start of muon chamber installation in ATLAS, 88 chambers
built at the Max Planck Institute for Physics in Munich have to be
commissioned and calibrated.  With a data taking period of one day 
individual wire
positions can be measured with an accuracy of $8.3\,\mu{\rm m}$ in the
chamber plane and $27\,\mu{\rm m}$ in the direction perpendicular to
that plane.
\end{abstract}
\end{titlepage}

\section{Introduction}

\begin{figure}[ht!]
\begin{center}
        \includegraphics[width=0.6\linewidth]{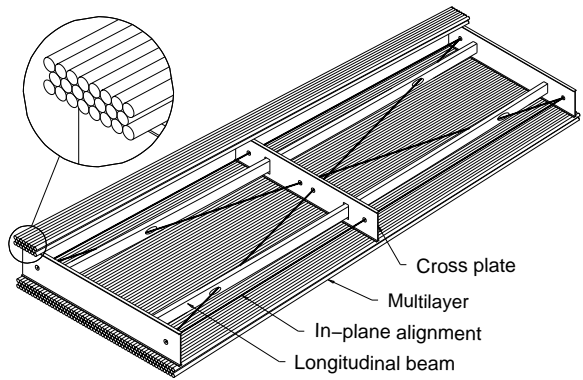}
        \caption{\label{mdt.fig} 
          A schematic view of an ATLAS MDT chamber. Also indicated
          are the optical paths of the geometry monitoring system
          (in-plane alignment).}
\end{center}
\end{figure}

The large majority of the precision detectors of the muon spectrometer
of the ATLAS experiment at the Large Hadron Collider (LHC) will be
Monitored Drift Tube (MDT) chambers. These chambers consist of 2
multilayers, each built of 3 or 4 layers of densely packed drift tubes
mounted on a support frame made of aluminum (compare
figure~\ref{mdt.fig}). The drift tubes --- which are also made of
aluminum --- have an outer diameter of 3 cm and a wall thickness of
400 $\mu$m; in the middle of each tube a gold-plated W-Re anode wire
of 50 $\mu$m diameter is stretched. The anode wire is positioned only
at the two tube ends by precision end-plugs; the sag of each chamber
can be adjusted to follow the sag of the wire along the tube. The
drift tubes are operated at a pressure of 3 bar absolute with an
Ar:CO$_{2}$=93:7 gas mixture at a gas gain of $2\times 10^{4}$. The
area covered by a complete chamber ranges from $1\,\mathrm{m}^2$ to
$11\,\mathrm{m}^2$. A total of 1194 MDT chambers will be used in
ATLAS~\cite{atlasmuontdr}.

In collaboration with the Max Planck Institute (MPI) for Physics,
Munich University (LMU) is responsible for the construction of 88
MDT chambers of type BOS (Barrel Outer Small --- the second largest type
used in the ATLAS barrel region) which consists of 432 drift tubes,
arranged in $2\times 3$ layers of 72 tubes, and has a
size\footnote{Some 
  chambers will be narrower or have
  cut-outs --- realized with shorter drift tubes --- to accommodate support
  structures of the ATLAS detector or feedthroughs to inner
  components.}
of $3.9\,\mathrm{m} \times 2.2\,\mathrm{m} \times 0.5\,\mathrm{m}$. 
The chambers are built with high mechanical precision
at MPI~\cite{mpipaper} and are then commissioned and calibrated at the
Cosmic Ray Measurement Facility of Munich University.

The main objectives of the Cosmic Ray Facility are to test the
chambers, to map the positions of the tube layers and of individual
anode wires within the chambers, and to gain experience in the
operation of the chambers and the calibration procedures.

An important design goal of the ATLAS muon spectrometer is the standalone
measurement capability of muon momenta up to 1\,TeV with a
relative error of less than 10\%. This leads to the requirement that
the anode wire positions within the MDT chambers must be known to within 
$20\,\mu{\rm m}$~\cite{atlasmuontdr}.

\section{The Cosmic Ray Measurement Facility Setup}
\begin{figure}[ht!]
\begin{center}
        \includegraphics[width=\linewidth]{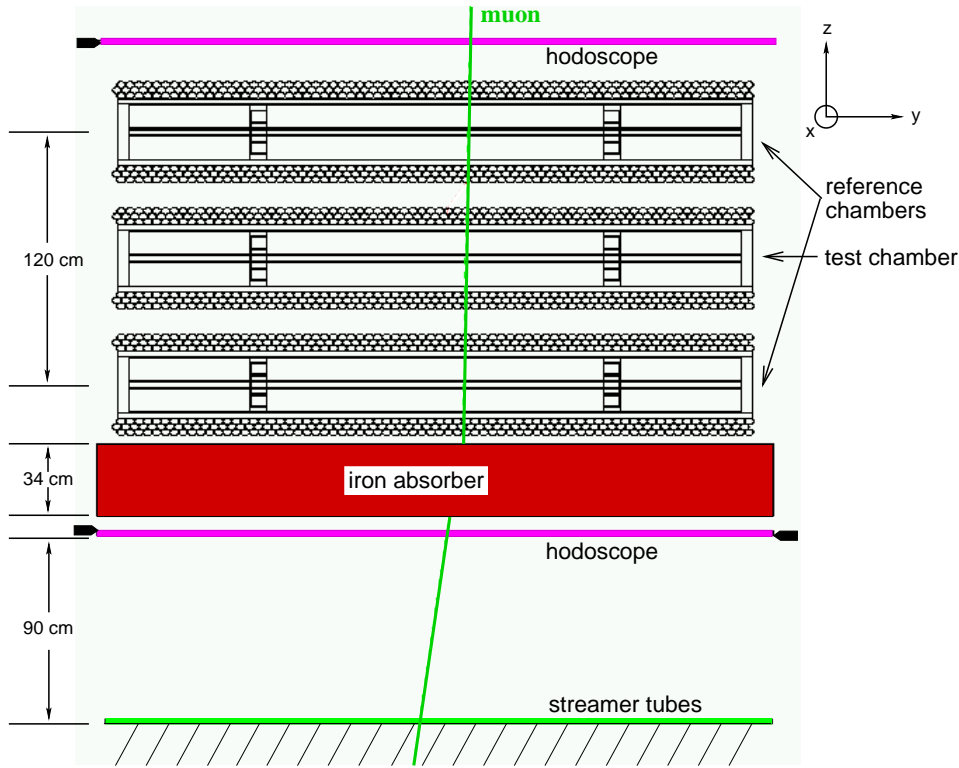}
        \caption{\label{fig:setup} A schematic view of the Cosmic Ray
        Measurement Facility.  The individual components are described
        in the text.}
\end{center}
\end{figure}
A schematic view of the Cosmic Ray Measurement Facility is shown in
figure~\ref{fig:setup}.  The whole setup is located in a climatized
hall to ensure stable environmental conditions during the
measurements.  We define a right handed coordinate system, the $x$
coordinate in the direction along the wires, $y$ perpendicular to the
wires in the chamber plane, and $z$ perpendicular to the chamber plane
pointing upwards.

Three MDT chambers are operated simultaneously in the facility. Two of
them, the so-called reference chambers, encompass the third
chamber --- which is to be tested --- and provide track information for 
cosmic ray muons.
The anode wire positions of the reference chambers have been mapped
with a precision of $2\,\mu\mathrm{m}$ with an X-ray tomograph at 
CERN~\cite{xray-tomograph}.

A coincidence of the three scintillator layers 
in the hodoscopes above and below the
MDT chambers provides the trigger. The upper hodoscope consists of one
layer of scintillator counters, the lower of two staggered layers
displaced by half the width of a counter.
Only muons with a momentum greater than 600\,MeV pass
the 34\,cm iron absorber above the lower hodoscope and can generate a
trigger signal.  The trigger logic is divided into 5 segments along
the $x$ coordinate to limit the inclination of the muons in the
$x$-$z$ plane.  The arrangement of the scintillator bars allows the
measurement of muon tracks in the $x$ direction with a precision of
8\,cm. The time of the muon transition is determined by the lower
hodoscope with a precision of better than 800\,ps. The average trigger
rate is about 70\,Hz.

The streamer tubes at the bottom of the setup
are used together with the lower reference chamber to determine the
multiple scattering angle in the $y$-$z$ plane in the iron absorber. This
angle is used in combination with the angle between the track segments in the
two reference chambers to estimate the momentum of the muon.

Two contact free alignment systems are used to continuously monitor
the positions of all MDT chambers: The position of the reference
chambers relative to each other is measured with an optical alignment
system consisting of eight RasNik sensors~\cite{rasnik}, the position
of the test chamber with respect to the upper MDT chamber with eight
capacitive sensors~\cite{capacitec}. Both systems have a precision of
better than $5\,\mu\mathrm{m}$~\cite{dipl}.

In addition to the chamber-to-chamber alignment, the internal chamber
geometry is also monitored with integrated RasNik sensors 
(compare figure~\ref{mdt.fig}).

\section{Drift Time Spectra} 
\label{sec_dt_fit}

The position measurement in the MDT chambers is based on the
measurement of the drift time of the electrons (from ionization 
by the incident muon along its trajectory) to the anode wire. The drift
time is determined with respect to the external trigger system and
translated to the impact radius (distance from the anode wire)
via the so-called $r$-$t$ relation (see
section~\ref{rt-rel-sec}); the drift tubes are only sensitive to the
track coordinates in a plane perpendicular to the anode wire, not
along it. The drift of the electrons depends on the gas composition,
the gas density, and the electric field applied (i.e.\ the operating 
voltage), and is therefore characteristic of the gas mixture and the
operating conditions.

After correcting for the signal propagation times along the anode wire
and in the electronics and cables, the distributions of drift times are used
as a first check that each tube is functioning as expected and that
the detector response is homogeneous.  For the simple case of 
uniform illumination of a
tube, the density of hits at a given drift time is proportional to
the drift velocity at the corresponding radius.

Events with more than one charged particle passing through the setup
would lead to ambiguities in the hodoscope time measurement.
Therefore, exactly one hit in the upper hodoscope and one
pair of hits in overlapping counters in the lower hodoscope are
demanded, which rejects 15\,\% of all triggers (see
table~\ref{cutflowtable1.tab}).

\begin{figure}[ht!]
\begin{center}
        \includegraphics[width=\linewidth]{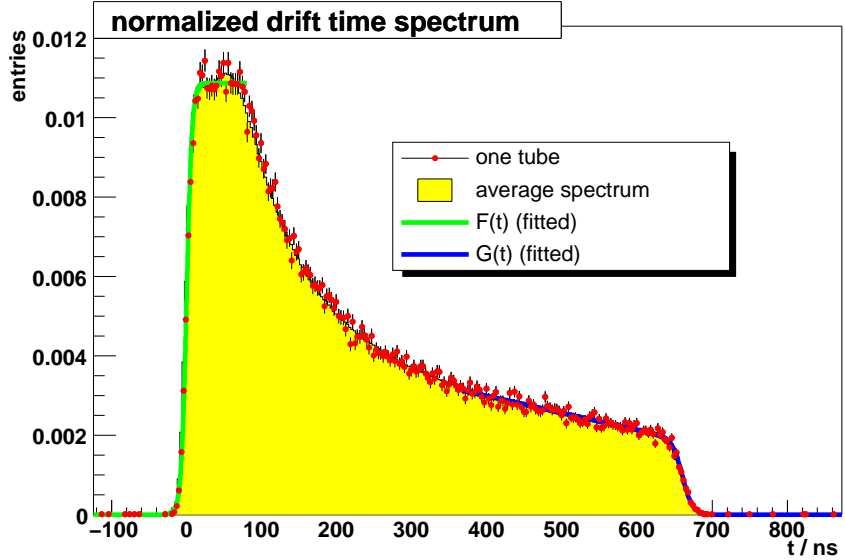}
        \vspace{-250pt}
        \caption{\label{fermi_fit} 
        A typical drift time spectrum, normalized to one entry.  The fitted
        functions are explained in the text.} 
\end{center}
\end{figure}

The edges of the drift time spectra are
parameterized~\cite{t0fit} with the functions
\begin{eqnarray}
        F\left(t\right) = p_0 + \frac{A_0}{1+e^{\frac{t_0-t}{T_0}}}
     && {\rm for\ the\ rising\ edge\ and}\\
        G\left(t\right) = p_m + \frac{\alpha_m t + A_m}{1 + e^{\frac{t - 
        t_m}{T_m}}}
     && {\rm for\ the\ trailing\ edge\ of\ the\ spectrum,}
\end{eqnarray}
as shown in figure~\ref{fermi_fit}.  The parameters $p_0$ and $p_m$
correspond to the rates of accidental hits, $A_0$ to the height of the
spectrum, and $t_0$ to the mid-time and $T_0$ to the steepness of the
rising edge of the spectrum, which 
is determined by the resolution of the drift tube near the anode wire.
Correspondingly, $t_m$ is the
middle of the trailing edge of the spectrum and $T_m$ its steepness,
which is determined by the decrease in pulse height near the tube 
wall\footnote{The small pulse height for muons passing a tube
  close to the tube wall is due to their short
  path inside the tube, which leads 
  to smaller primary ionization.}.  
The parameterization with $\alpha_m t + A_m$ accounts
for the slope of the spectrum before its trailing edge that is due to
the decrease of the drift velocity with increasing radius.

\begin{figure}[ht!]
\begin{center}
        \includegraphics[width=\linewidth]{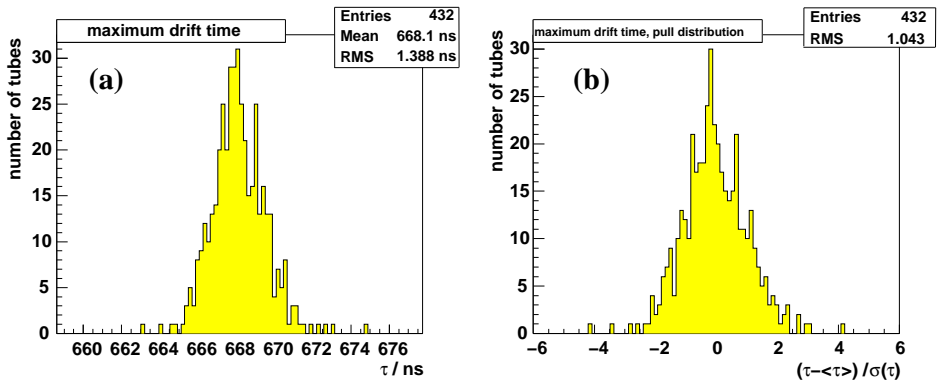}
        \caption{\label{dt_dis} a) Distribution of the maximum drift
          time $\tau=(t_m+T_m) - t_0$ for the tubes of a tested chamber. b)
          The corresponding pull distribution.}
\end{center}
\end{figure}

The maximum drift time, defined as $\tau = (t_m + T_m) - t_0$, is
characteristic of the drift properties of a tube.  The 
distribution of the maximum drift time is shown in figure~\ref{dt_dis}a
for the tubes of one MDT chamber.
As part of the chamber quality control programme, it is checked that 
all tubes have the same drift time spectrum characteristics within errors,
compare figure~\ref{dt_dis}b. 

\section{The Relation between Drift Time and Drift Radius}
\label{rt-rel-sec}

The relation between the measured drift time and the corresponding
drift radius (\mbox{$r$-$t$} relation) is determined without the help of external
spacial detectors and without assumptions on uniform tube
illumination.  
After the check described in section~\ref{sec_dt_fit}, 
it is assumed that all tubes of an MDT chamber have the
same $r$-$t$ relation.

Events with two hits at small drift times ($|t-t_0|<15\,{\rm ns}$)
in one MDT chamber 
are selected. By assuming that the muon went through 
the two corresponding anode wires and that the
trajectory of the muon is a straight line, 
the impact radii in the other tubes of the chamber can be
calculated.
This procedure yields drift times for a discrete set of drift radii.
The number of such
points --- typically 16 --- is determined by the geometry of the chamber 
and the angular
acceptance of the trigger.  
In addition to these points, the drift times corresponding to $r=0$ and to the
inner tube radius are determined by
the fit to the drift time spectra described in
section~\ref{sec_dt_fit}.  
The $r$-$t$ relation is then approximated by a linear interpolation between
the measured values.

\section{Track Reconstruction} 

Within each MDT chamber a muon track segment is reconstructed from
the measured hits. 
A track segment is assumed to be a straight line in the $y-z$ plane\footnote{The
third coordinate ($x$) is 
only used to correct for signal propagation times and for the wire sag
and is taken from the hodoscope information.}:
\begin{equation}
\label{track.eqn}
y(z) = m_n z + b_n,
\end{equation}
where the index $n$ specifies the chamber.
The parameters $m_n$ and $b_n$ are
determined for each chamber by minimizing the $\chi^2$ function
\begin{equation}
\chi^2 = \sum_i \frac{\left(r_{track,\,i} - r_{drift,\,i}\right)^2}
                     {\sigma_{i}(r_{drift,\,i})^2},
\end{equation}
where the index $i$ runs over all tubes contributing to the track segment,
$r_{track,\,i}$ is the distance of the track segment to the 
wire, $r_{drift,\,i}$ is the
measured drift radius, and $\sigma_{i}$ is the resolution of the tube
for the given radius.
At a test beam measurement using an external reference tracking
system of silicon strip detectors
the intrinsic single tube resolution was determined to be
around $100\,\mu{\rm m}$~\cite{x5paper}.

The tubes contributing to a track segment are selected by choosing 
one tube with a hit in each layer of the chamber, which is a sample
of six tubes.
From the impact radii in the two outermost tubes of this 
sample four possible track segment candidates are derived.
If the difference between the distance of such a candidate to
the wire and the measured drift radius does not exceed 2\,mm in any of
the other tubes, a track segment is fitted using this sample. If more than
one track segment is found, the one with the smallest $\chi^2$ is used.  If
no track segment is found, the procedure is repeated with subsamples of five
tubes.

\section{Chamber Alignment}
\label{align-sec}

The MDT chambers in the Cosmic Ray Facility are aligned mechanically
with a precision of about $100\,\mu$m.  A much more precise estimate of
their relative positions is then obtained by comparing the muon track segments
reconstructed in each chamber.  The optical and capacitive alignment
systems are used to verify that the setup does not move significantly
during the data taking period.
In the future, the alignment information will be included
in the analysis to correct for time-dependent position shifts of the chambers.

The positions of the two reference chambers are
measured relative to the test chamber.
\begin{list}{$\bullet$}{\setlength{\itemsep}{0ex}
                        \setlength{\parsep}{0ex}
                        \setlength{\topsep}{0ex}}
\item
A shift $v_y$ of a reference chamber with respect to
the test chamber results in a systematic shift between the track segments
in the reference and test chamber and can be obtained from the average
over all events,
\begin{equation}
        v_y = \left< b_{\rm ref} - b_{\rm test}\right> \ ,
\end{equation}
where $b_{\rm ref}$ and $b_{\rm test}$ are the track segment parameters
defined in equation~(\ref{track.eqn}). They are determined with a
precision of 1\,$\mu$m.

\item
A shift $v_z$ results in a relative shift of the track segments depending on
the slope $m$. It is determined from a linear fit to the
distribution of $b_{\rm ref} - b_{\rm test}$ versus $m$ with an error
of 10\,$\mu$m. 

\item
A shift $v_{x}$ along the $x$ coordinate can be ignored, as the MDT
chambers are not sensitive to such a shift along the direction of
their anode wires.

\item
The tilt angle $\alpha$ of a reference chamber around the $x$ axis is given
by the systematic deviation of the two track segment slopes as
\begin{equation}
\alpha = \left< m_{\rm ref} - m_{\rm test}\right> \ .
\end{equation}
A precision of $10^{-6}$ is achieved.

\item
To determine the angles $\beta$ and $\gamma$ corresponding to
rotations of the reference chamber around the $y$ and $z$ axes, the
$x$ coordinate measurement of the hodoscope is used. The shifts $v_y$
and $v_z$ are calculated for three one meter wide sections in $x$.
From the dependences $v_z(x)$ and $v_y(x)$, the angles $\beta$ and
$\gamma$ are obtained with precisions of $8\times10^{-6}$ and $10^{-6}$, 
respectively.
\end{list}
\vspace{1ex}

In the further analysis, 
the parameters of track segments in the reference chambers 
are corrected according to the measured reference chamber positions.

\section{Energy Estimation}
\label{energy-sec}

The energy of cosmic muons traversing the Cosmic Ray Measurement
Facility is estimated from multiple scattering --- which depends on the
muon momentum --- in the chambers and the iron absorber.

If $m_{\rm ref,\,u}$ designates the slope of the track segment reconstructed
in the upper reference chamber and $m_{\rm ref,\,l}$ the slope of the
track segment reconstructed by the lower reference chamber, the distribution
of $\Delta m = m_{\rm ref,\,u}-m_{\rm ref,\,l}$ will be the wider the
lower the muon energy.  The standard deviation $\sigma_{\Delta m}(E_\mu)$ 
has been determined as a function of the muon energy by
means of a Monte Carlo simulation and can be parameterized
as~\cite{kortner}
\begin{eqnarray}
\sigma_{\Delta m}(E_\mu) = 
\sigma_\infty + \sigma_0 \left(\frac{600\,{\rm MeV}}{E_\mu}\right)^\alpha \ ,
\end{eqnarray}
with $\sigma_\infty=(2.0\pm0.2)\cdot10^{-4}$, $\sigma_0 = (1.001\pm0.008)\cdot
10^{-2}$, and $\alpha=1.04\pm0.01$.

Similarly, the deviation $\Delta y$ of the position measured with the 
streamer tubes from the prediction obtained from the track segment in the lower
reference chamber depends on the muon energy: 
For its width $\sigma_{\Delta y}(E_\mu)$, a Monte
Carlo simulation yields the values
$\sigma_\infty=(4.4\pm0.1)$~mm, $\sigma_0=(183\pm2)$~mm, and
$\alpha=1.44\pm0.01$.

The probability density $f$ for measuring deviations $\Delta m$ and $\Delta y$ 
is then given by
\begin{eqnarray}
f(E_\mu) =
\frac{1}{2\pi} 
\frac{1}{\sigma_{\Delta m}(E_\mu)} 
\frac{1}{\sigma_{\Delta y}(E_\mu)} 
\exp\!\left(  -\frac{1}{2}
              \left[     \left(\frac{\Delta m}{\sigma_{\Delta m}(E_\mu)}\right)^2
                    \!+\!\left(\frac{\Delta y}{\sigma_{\Delta y}(E_\mu)}\right)^2
              \right] 
      \right) \ .
\end{eqnarray}
For a given track, the value of $E_\mu$ that maximizes 
$f(E_\mu)$ is an estimate of the muon energy.
This estimator
is biased and has a limited resolution 
but is still useful for a selection of high momentum muons.

\section{Measurement of Wire Positions}

In this section, the 
method for the measurement of anode wire positions is presented.  To
determine wire positions, the muon track segments in the reference chambers
are extrapolated into the test chamber, and the drift radii measured
in the test chamber are compared with the track predictions.

The Cosmic Ray Measurement Facility is particularly sensitive to
displacements $\delta y$ of the wires in the chamber plane. From this,
a shift of
entire tube layers in the $y$ direction and the mean spacing $g$
between the wires within a layer can also be determined.

Wire displacements $\delta z$ in the plane perpendicular to the
chamber are accessible via tracks with different inclinations $m$
(cf.~equation~(\ref{track.eqn})).  This measurement is less precise than
the $\delta y$ determination owing to the limited angular
acceptance of the trigger and the angular distribution of cosmic
muons.   
During chamber production, $z$ displacement and tilts of entire tube
layers are more difficult to control than individual wire
displacements $\delta z$ with respect to the
layer.  Therefore, the $z$
displacement for each tube layer is also measured.

The measurements are performed at both ends of the chamber 
by selecting muons which passed the tested chamber within one meter
of the corresponding end of the chamber using the hodoscope information.
Because the wires are only supported at the tube ends and since
the sag of the anode wires is known from wire tension measurements
these two measurements completely determine the wire position.

\subsection{Event Selection}

Events with one reconstructed charged particle are selected as described in 
section~\ref{sec_dt_fit}.  For the further analysis, 
the event selection proceeds in two steps: First, the reconstructed
track segments are subjected to a set of cuts, and second, the individual hits
in the tubes must pass certain quality criteria.

The presence of a reconstructed track segment with at least 5 hits
is required in each of the three MDT chambers.
The track segments must also roughly match, i.e., 
$|b_{\rm ref,\,u}-b_{\rm ref,\,l}|<4\,{\rm mm}$
and $|m_{\rm ref,\,u}-m_{\rm ref,\,l}|<15\,{\rm mrad}$. This ensures
that the track segments are not affected by tubes in which delta
rays\footnote{Energetic primary electrons knocked from the tube
  walls or gas atoms by the incident charged particle. Delta rays can
  lead to smaller measured drift times if they are emitted in the
  direction of the anode wire.} 
distorted the drift time measurement. The remaining events are called
good tracks.
For the determination of wire positions, tracks with high muon momentum are
desirable to limit the effects from multiple scattering.  Thus, the estimated
muon energy is required to be larger than $2.5\,{\rm GeV}$.
The efficiencies of these track cuts are given in table~\ref{cutflowtable1.tab}.

Tubes in the test chamber with hits generated by delta rays would affect the
measurement of the wire position, and are therefore rejected: It is
demanded that the tube have contributed to a track segment in the test
chamber. 
Also, a loose cut on the difference between the radius prediction from
the reconstructed track segments and the measured drift radius is made. For
the track segment reconstructed in the test chamber this cut is at 0.7\,mm,
while it is at 1.5\,mm for the predicted radius from the reference
track segments extrapolated into the test chamber\footnote{The larger value for
  the reference chamber track segments 
  is due to the long extrapolation distance
  into the test chamber.}. The efficiencies of these
cuts are given in table~\ref{cutflowtable2.tab}. They are
very close to 100\% and
thus do not bias the wire position measurement.

\subsection{Determination of the Wire Positions}
\label{position.subsec}

The reconstructed reference chamber track segments 
are extrapolated into the test
chamber and the drift radius measured in the test chamber is
compared with the prediction from those track segments.
The weighted average $r_{\rm ref}$ of the drift radius predictions from the 
two track segments
\begin{equation}
\label{rref.eqn}
r_{\rm ref} = 
\frac{   (1/\sigma^2_{\rm u}) r_{\rm ref,\,u}
       + (1/\sigma^2_{\rm l}) r_{\rm ref,\,l}}
     {   (1/\sigma^2_{\rm u})
       + (1/\sigma^2_{\rm l})}
\end{equation}
is taken, where each uncertainty $\sigma_i$ ($i={\rm u,\,l}$ for the
track segment in the upper and lower reference chamber) depends 
on the extrapolation
distance from the reference chamber and the estimated muon momentum
(to account for the uncertainty from multiple scattering). The value
of $\sigma_i$ is about $200\ \mu$m for the reference chamber closer to
the tube hit and ranges from 350 to 800 $\mu$m for the farther chamber.

\begin{figure}[t!]
\begin{center}
        \includegraphics[width=0.5\linewidth]{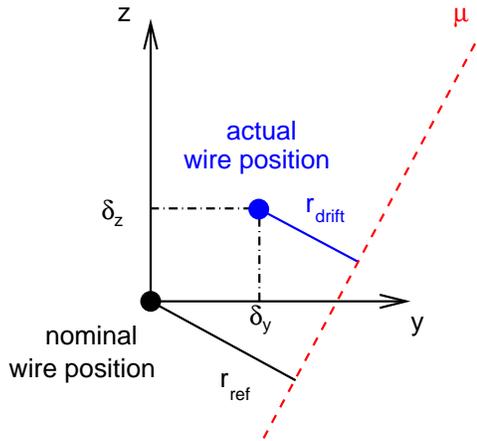}
        \caption{\label{deltarsketch.fig} A sketch explaining the method for 
          the measurement of wire displacements.  A muon (dashed line)
          passing a tube with a wire that is displaced from the
          nominal position by $\delta y$ in the $y$ direction and $\delta
          z$ in the $z$ direction will lead to a measured drift radius
          $r_{\rm drift}$, while $r_{\rm ref}$ is predicted by the
          reference chambers relative to the nominal wire position.
        }
\end{center}
\end{figure}
If a given wire is displaced by $\delta y$ in the $y$ direction and
$\delta z$ in the $z$ direction, the measured drift radius $r_{\rm drift}$
in the tube will show a systematic deviation $\Delta r$ from the
predicted reference radius,
\begin{equation}
\label{deltar.eqn} 
\Delta r = r_{\rm drift} - r_{\rm ref} = \delta y - m \, \delta z + {\cal O}(m^2) \ ,
\end{equation}
as shown in figure~\ref{deltarsketch.fig}.

\subsection{Determination of the Wire Position Perpendicular to the
  Chamber Plane}
\label{zposition.subsec}

The displacement $\delta z$ of an individual anode wire is 
determined from the slope in a linear fit to the distribution of
$\Delta r$ vs.~$m$.

The $z$ displacement of an entire tube layer and its 
tilt corresponding to a rotation about the
$x$ axis are determined from the distribution of
measured $\delta z$ values of all tubes as a function of their $y$
coordinate\footnote{Here, $y$ is taken to be the nominal wire
  position. Because of the large chamber width, this approximation 
  does not bias the result.}.
A linear fit to this $\delta z$ distribution is
performed separately for each layer.  The offset and slope of the
fitted line are taken to be the $z$ displacement and tilt of the
layer, respectively.

In the further analysis, each wire is assigned the measured $z$ position of the
layer at the nominal $y$ wire position.  
The single wire position measurement is used for wires where it
deviates from the layer
position by more than 3 standard deviations.

\subsection{Determination of the Wire Position in the Chamber Plane}
\label{yposition.subsec}

Using the $z$ position measurement described above, 
the displacement $\delta y$ of an anode wire
is obtained as
\begin{equation}
\label{deltaraverage.eqn}
\langle\Delta r'\rangle = \langle r'_{\rm ref} - r_{\rm drift} \rangle
= \delta y - \langle m \rangle \delta z' \approx \delta y \ ,
\end{equation}
where $r'_{\rm ref}$ denotes the predicted radius corrected for the
wire $z$ position and the residual displacement $\delta z'$ after
correction is of the order of the resolution of the $z$ position
measurement.
The average of the track slopes
$\langle\left|\langle m\rangle_{\rm all\,hits}\right|\rangle_{\rm all\,tubes}$
is\footnote{The deviation from 0 is caused by a slight
  asymmetry of the trigger efficiency in the $y$ direction during data
  taking.} 
$0.06$,
the largest value of $\left|\langle m\rangle_{\rm all\,hits}\right|$
for a single tube being $0.15$.  Therefore, the systematic
error introduced in the last approximation of 
equation~(\ref{deltaraverage.eqn})
is typically much smaller than $15\%$ times the resolution on the $z$ position,
which does not limit the $\delta y$ measurement.

In the chamber plane, the nominal wire positions can be described by a
grid of the form
\begin{equation}
\label{grid-eq}
y(n) = y_0 + n \, g \ ,
\end{equation}
where $n$ denotes the index of the tube in the layer, $y_0$ the offset
of the layer position relative to the other layers in the same
chamber, and $g$ the mean distance between
two neighbouring wires.
Fitting the function in equation~(\ref{grid-eq}) to the distribution
of the wire positions of one tube layer yields the offset $y_{0}$ and
the mean distance $g$ for that layer. 

\subsection{Test of the Method}

In order to test the method described above, one of the few chambers
which have been scanned by the X-ray tomograph at CERN has been used
as a test chamber.  Two tomograph scans of the chamber were performed
at a distance of 30\,cm from each chamber end.  Therefore the
positions of the anode wires\footnote{Only
typically $80\%$ of the wires are visible in the tomograph scans because of 
the support structure of the chamber.} 
are known with a precision of 2\,$\mu$m.

The analysis uses 28 hours of data taken with this chamber.  During
this time 1.7~million events per meter along the anode wires
($x$ direction) have been recorded.  The number of events after each
step of the event selection and the number of tube hits in the test
chamber used for the wire position measurements are given in
tables~\ref{cutflowtable1.tab} and~\ref{cutflowtable2.tab}. 

All measurements at the Cosmic Ray Facility are in very good
agreement with the tomograph results.

\begin{table}
\begin{center}
\begin{tabular}{|l|c|c|}
\hline
selection cut & 
\begin{tabular}{@{}c@{}}number of\\ events after cut\end{tabular} &
\begin{tabular}{@{}c@{}}efficiency\\ of cut\end{tabular} \\
\hline
all events & $1.65\cdot10^6$ & \\
hodoscope hits from one particle & $1.40\cdot10^6$ & 85\,\% \\
good tracks & $0.84\cdot10^6$ & 60\,\% \\
estimated energy $>2.5\,{\rm GeV}$ & $0.34\cdot10^6$ & 40\,\% \\
\hline
\end{tabular}
\caption{\label{cutflowtable1.tab}
  Number of events per meter along the wire after each cut in the
  event selection (second column) and the percentage of
  events which pass the cut (right column).}
\end{center}
\end{table}

\begin{table}
\begin{center}
\begin{tabular}{|l|c|c|}
\hline
selection cut & 
\begin{tabular}{@{}c@{}}number of\\ tube hits after cut\end{tabular} &
\begin{tabular}{@{}c@{}}efficiency\\ of cut\end{tabular} \\
\hline
all tube hits & $2.04\cdot10^6$ & \\
hits on tracks & $1.90\cdot10^6$ & 96\,\% \\
residuum cuts & $1.87\cdot10^6$ & 98\,\% \\
\hline
\end{tabular}
\caption{\label{cutflowtable2.tab}
Number of tube hits in the test chamber per meter along 
the wire (as used for the wire position measurement) after each 
cut in the event selection (second column) and the percentage of
hits which pass the cut (right column).}
\end{center}
\end{table}

\begin{figure}[htp]
\begin{center}
        \includegraphics[width=\linewidth]{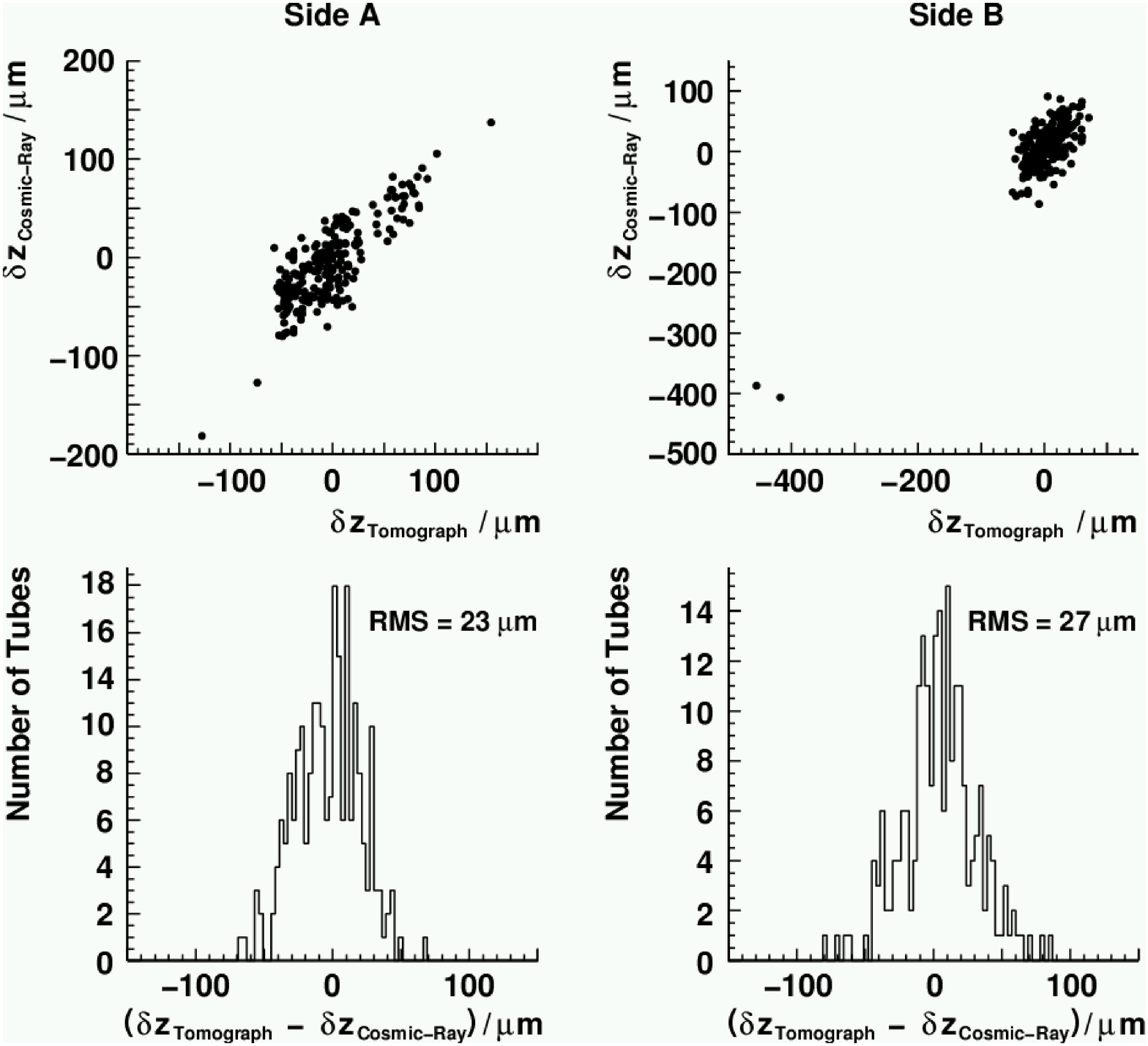}
        \caption{\label{z_dis} Comparison of the $\delta z$ measurement
        for single wires at the Cosmic Ray Facility with the tomograph scans.
	Note the different scales in the two upper plots.
}
\end{center}
\end{figure}
In figure~\ref{z_dis}, the $\delta z$ measurements for single wires
performed at the Cosmic Ray Facility are compared with the results of
the tomograph scans. The distributions of the difference between the
cosmic ray measurement and the tomograph scan have widths of 23~$\mu$m
and 27~$\mu$m.  These widths are dominated by the resolution of 
the Cosmic Ray Measurement Facility.

\begin{figure}[htp]
\begin{center}
        \includegraphics[width=\linewidth]{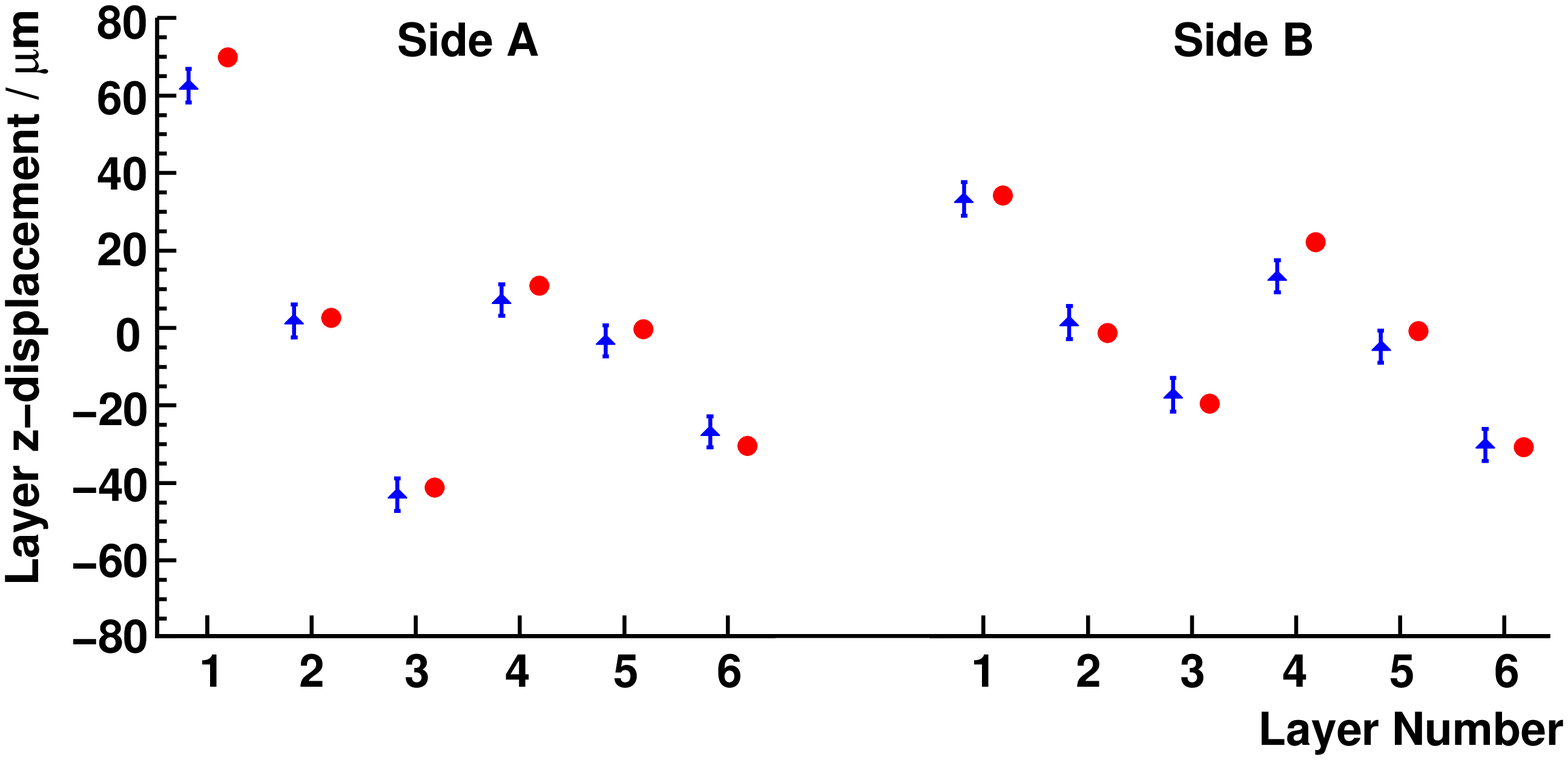}
        \caption{\label{lz_pos} Measurements of $\delta z$ for entire
        tube layers. 
        The triangles denote the cosmic ray measurement, the 
        dots the X-ray tomograph data.}
\end{center}
\end{figure}
\begin{figure}[htp]
\begin{center}
        \includegraphics[width=\linewidth]{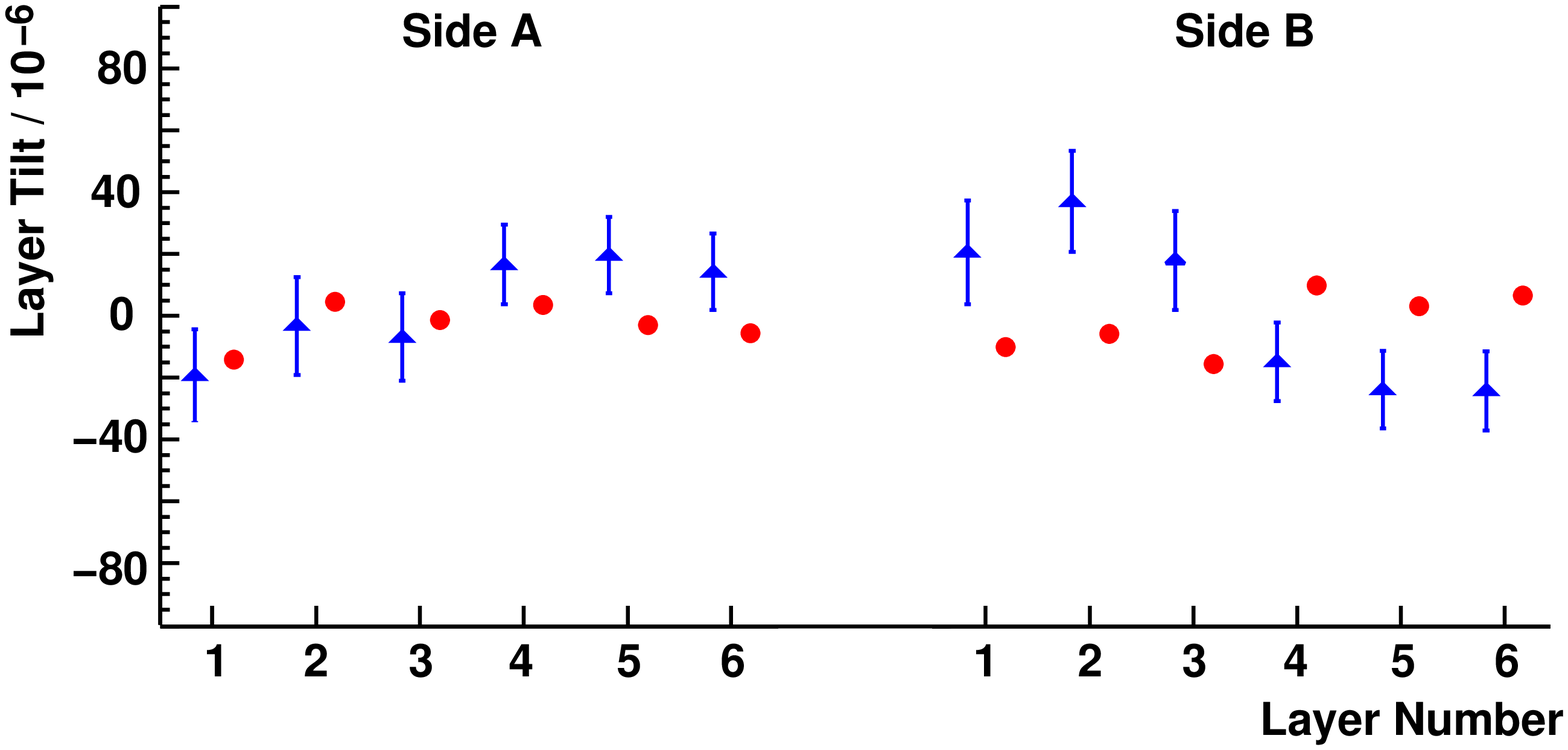}
        \caption{\label{lz_slope} Measurements of the tilt of layers.
         The triangles denote the cosmic ray measurement, the 
         dots the X-ray tomograph data.}
\end{center}
\end{figure}
A better precision can be obtained for the $z$ displacement 
of entire tube layers.
A comparison of this measurement and the 
values for the tilt angle of the tube layers around the $x$ axis 
(the parameters
which are more difficult to control during the MDT chamber assembly)
with the tomograph results is shown in figures~\ref{lz_pos} 
and~\ref{lz_slope}.
Here, we achieve a precision of $4.4\,\mu{\rm m}$
for the layer shift
and $17\times10^{-6}$ for the tilt angle.

\begin{figure}[tp]
\begin{center}
        \includegraphics[width=\linewidth]{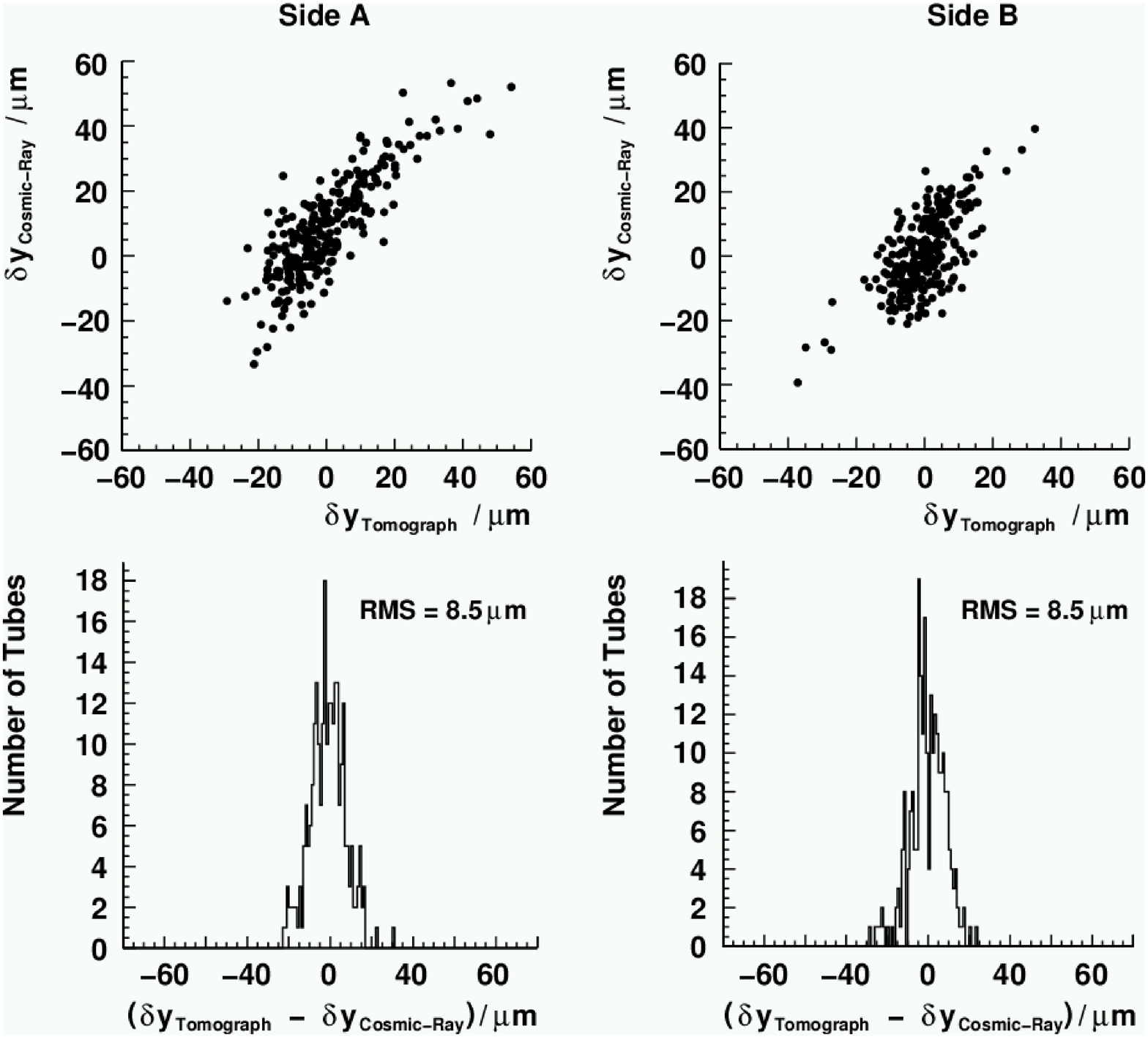}
        \caption{\label{y_dis} Comparison of the $\delta y$ measurement
        for single wires at the Cosmic Ray Facility with the tomograph
        scans.}
\end{center}
\end{figure}
The comparison of the $\delta y$ measurement with cosmic rays with the
tomograph results shows that the difference between the two
measurements has an RMS of $8.5\,\mu{\rm m}$ 
at either end
of the chamber (see figure~\ref{y_dis}). 
The precision of the 
Cosmic Ray Measurement Facility is therefore 8.3\,$\mu$m
(after subtraction of the uncertainty on the X-ray tomograph measurement).

\begin{figure}[htp]
\begin{center}
        \includegraphics[width=\linewidth]{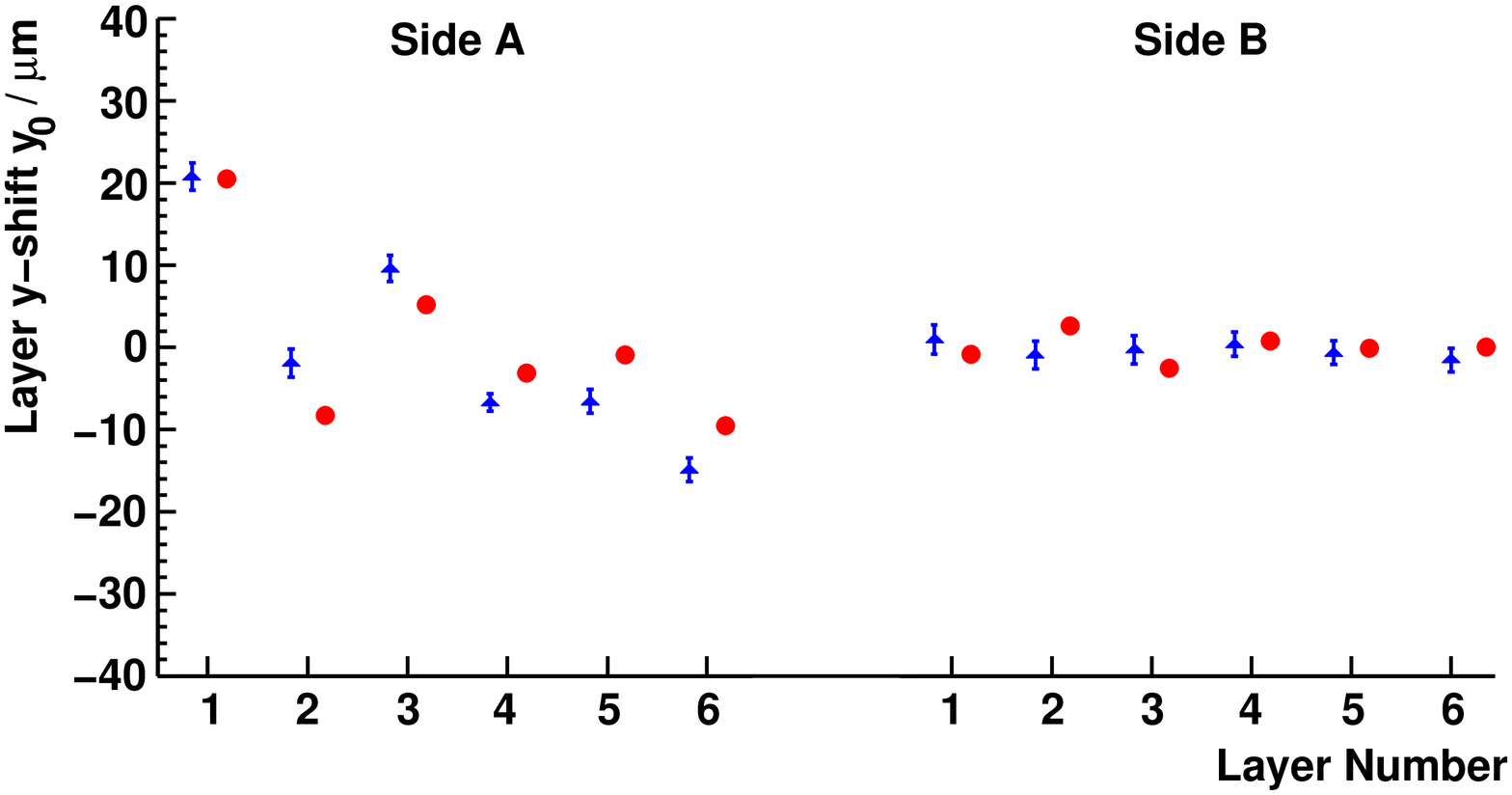}
        \caption{\label{yshift.fig} Measurements of the layer shift
        in $y$.  
        The triangles denote the cosmic ray measurement, the 
        dots the X-ray tomograph data.}
\end{center}
\end{figure}
\begin{figure}[hpt]
\begin{center}
        \includegraphics[width=\linewidth]{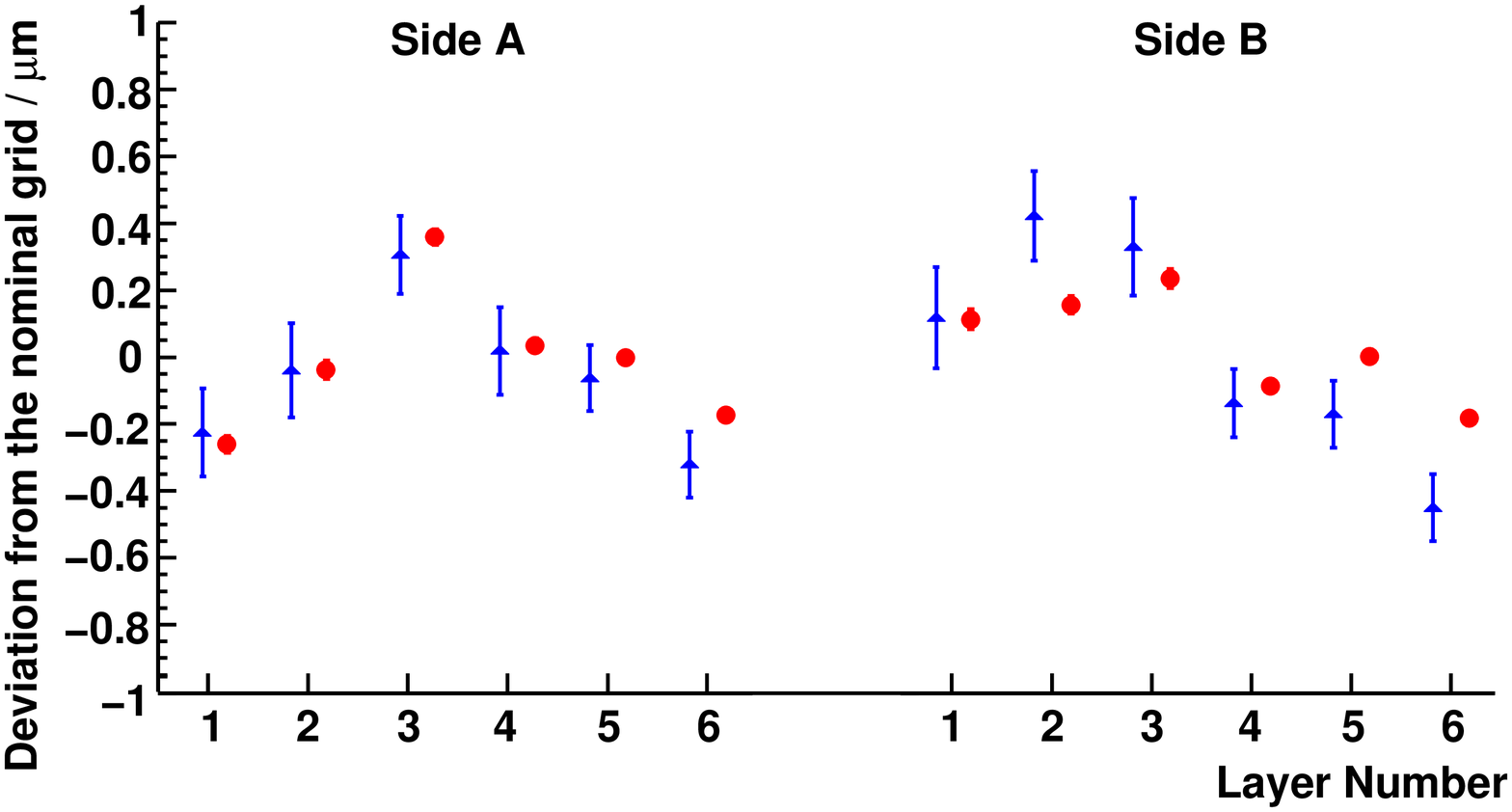}
        \caption{\label{ygrid.fig} Measurements of the deviation of the
        grid from the nominal grid.  
        The triangles denote the cosmic ray measurement, the 
        dots the X-ray tomograph data.}
\end{center}
\end{figure}
The measurements of the layer offset $y_0$ and the wire grid constant
$g$ are compared in figures~\ref{yshift.fig} and~\ref{ygrid.fig}.  
A precision of $1.8\,\mu{\rm m}$ for the layer offset 
and $0.15\,\mu{\rm m}$ for the
grid constant is achieved.

\section{Conclusions}
\label{results.sec}

In the Cosmic Ray Measurement Facility, ATLAS MDT muon chambers are
tested and calibrated.  The Cosmic Ray Measurement
Facility is capable of measuring 
the positions of the individual anode
wires of a chamber with a precision 
of $8.3\,\mu{\rm m}$ in the
chamber plane and $27\,\mu{\rm m}$ in the
direction perpendicular to that plane 
after 28 hours of data taking.

In addition to the wire positions, the offsets of the tube
layers relative to each other in the chamber plane and their grid
constants have been determined with precisions of 
$1.8\,\mu{\rm m}$ and
$0.15\,\mu{\rm m}$, respectively.
The positions of the tube layers in the direction 
perpendicular to the chamber plane and the angles for
a rotation around
an axis parallel to the wires have been measured with precisions of 
$4.4\,\mu{\rm m}$ and 
$17\times10^{-6}$, respectively.
 
At a rate of up to two tested chambers per week, all 88 BOS chambers
which are produced in Munich will be tested until the planned start of muon
chamber installation in ATLAS.

\section{Acknowledgements}

We would like to thank our colleagues at Dubna and MPI Munich for the 
excellent collaboration, the X-ray tomograph group at CERN for providing
their results, and the staff at LMU for their support.
We are pleased to acknowledge the support of the Maier-Leibnitz-Laboratorium
of LMU and TU Munich, and the Bundesministerium f\"ur Bildung und 
Forschung, Germany.


\begin{thebibliography}{5}
\bibitem{atlasmuontdr} 
The ATLAS Muon Collaboration, 
{\em ATLAS Muon Spectrometer Technical Design Report,}
CERN/LHCC 97-22, June 1997.

\bibitem{mpipaper} F.~Bauer et al., MPI Report, MPI-PhE/2002-04,
October 2002;\\
F.~Bauer et al., Nucl.~Instr.~and~Meth.\ A 461 (2001) 17;\\
F.~Bauer et al., IEEE Trans.~Nucl.~Sci.~48 (2001) 302.

\bibitem{xray-tomograph} J.~Berberis et al., {\em High-precision 
X-ray tomograph for
quality control of the ATLAS muon monitored drift chamber,} Nucl. Instrum.
Methods Phys. Res.~A~419~(1998)~342-350.

\bibitem{rasnik} H.~van der Graaf, H.~Groenstege, F.~Linde, and
P.~Rewiersma, {\em RasNiK, an Alignment System for the ATLAS MDT Barrel Muon
Chambers - Technical System Description,} NIKHEF/ET38110, 2000.

\bibitem{capacitec} S.~W.~Mackall, {\em Measurement of the Stability in the
Relative Alignment between the Silicon Microvertex Detector and the Time
Expansion Chamber Subdetectors in the L3 Experiment at CERN during 1994 Large
Electron Positron Collider Run,} Master thesis, Tuscaloosa, 1995.

\bibitem{dipl} A.~Kraus, {\em Genaue Bestimmung der Ereigniszeit und
Entwicklung eines Alignierungssystems f\"ur einen gro{\ss}en
H\"ohenstrahlteststand,} Diploma thesis, LMU Munich, 2001;\\ 
W.~Stiller, {\em Optical and Capacitive Alignment of ATLAS
Muon Chambers for Calibration with Cosmic Rays,} Diploma thesis, LMU
Munich, 2002.

\bibitem{t0fit} O.~Kortner and F.~Rauscher, {\em Automatic Synchronization of 
Drift-Time Spectra and Maximum Drift-Time Measurement of an MDT,} 
ATLAS internal note ATL-COM-MUON-2002-006, CERN, 2002.

\bibitem{x5paper} 
M.~Deile et al.,
{\em Resolution and 
Efficiency Studies with a BOS Monitored Drift-Tube Chamber and a Silicon 
Telescope at the Gamma Irradiation Facility,} ATLAS internal note,
ATL-COM-MUON-2003-006, CERN, 2003.

\bibitem{kortner} O.~Kortner, {\em Schauerproduktion 
durch hochenergetische Myonen 
und Aufbau eines H\"ohenstrahlungspr\"ufstandes f\"ur hochaufl\"osende
ATLAS-Myonkammern,} PhD Thesis, LMU Munich, 2002.

\end{thebibliography}
\end{document}